\documentclass[12pt]{article}  
\usepackage[dvips]{graphicx}
\usepackage{amsmath,amssymb,amsbsy}
\usepackage{amsfonts}
\usepackage{epsfig}
\newcommand{\beq}[1]{ \begin{equation}\label{#1}}
\newcommand{\eeq}{\end{equation}}
\newcommand{\beqa}{\begin{eqnarray}}
\newcommand{\eeqa}{\end{eqnarray}}

\newcommand{\no}{\nonumber}
\begin{document}\thispagestyle{empty}
\null
\begin{center}
\vskip 1.cm
{\Large\bf{Parton Distribution Functions properties\\
\vskip 0.35cm
of  the statistical model\\
}}
\vskip 1.cm
{\bf Claude Bourrely}\footnote{Electronic address: claude.bourrely@univ-amu.fr}
\vskip 0.3cm
 Aix-Marseille Universit\'e,
D\'epartement de Physique,\\ Facult\'e des Sciences, site de Luminy,
13288 Marseille, Cedex 9, France\\
\vskip 0.5cm
{\bf Abstract}
\end{center}
We show that the parton distribution functions (PDF) described
by the statistical model have very interesting physical properties
which help to understand the structure of partons.
The role of the quark helicity 
components is emphasized as they represent the building blocks of the
PDF.
In the model the sign of the  polarized quarks PDF comes 
out in a quite natural way once the thermodynamical potentials with a given 
helicity are known. Introducing the concept of entropy we study the states made 
of $|2u + d >$, $|u +d +s >$
and $|2\bar u +\bar d >$, for a fixed $Q^2$, the variation with $x$ shows that 
the first state has a dominant entropy due to the effect of  $u$ quark.
We prove that the PDF parameters obtained from experiments give in fact an 
optimal solution of an entropy equation subject to constraints. The same optimal
property is proven for the structure functions $F^2_p$ and $g^1_p$, and finally
to the quarks themselves.
 We develop a new approach of the polarized gluon density
based on a neural model which explains its property, in particular, a large
positivity value and an agreement with the positvity constraint.
An extension of this neural approach is applied to quarks giving
a coherent description of the partons structure.

\vskip 0.5cm

\clearpage
\newpage
\section{Introduction}
\setcounter{page}{1}
The role of the parton distribution functions in QCD theory
is essential to describe both unpolarized and polarized reactions, so a tremendous
effort has been undertaken to find the most accurate distributions. 
In the litterature
we remark that unpolarized and polarized PDF are treated as separated entities
\cite{lead2010}-\cite{delgado2013}, 
however, a simultaneous treatment of unpolarized and polarized PDF 
should in principle gives a more constrained determination.
Since many years we have adopted this point of
view in the framework of a statistical model where the PDF
 are built in from their helicity components.\cite{bou2002,bou2005}

In the absence of a theory for the parton distributions two approaches are
currently proposed. In one approach the distributions are approximated by
different polynomials which require numerous parameters, 
with such an hypercube a carefull numerical
analysis of the errors is necessary in order to get the most precise
values of the distributions, however
no attention is paid to the physical structure
of the partons nor on the meaning of the parameters values. In our approach
the physical structure of the distributions is introduced through a
statistical model, this information allows us to work with a reduce number
of parameters (21) whose meaning can be justified. In both case a good description
of the experimental data is obtained so we have a possible choice between
a numerical formulation of the distributions versus a physical one where
more emphasize is put on the partons structure \cite{bou2015b}.

The application of a statistical model, for instance, to a proton at rest 
which contains three quarks seems not justified due to the 
low number of elements. But when accelerated in a collider the energy
increase has not only an effect on its mass but also to create
a large number of $q~\bar q$ pairs or a quark gluon plasma
which in a p-p collison materialize mainly
in primary unstable particles observed in a detector as large number of tracks.
These occurence of numerous pairs provide 
a justification for a statistical treatment of the partons interaction process.

Let us mention an other application of the statistical model to different
elastic scattering reactions in terms of quarks PDF 
defined in impact parameter space, we have shown, in particular, 
that the gluon contribution
is essential to explain the dip structure of the pp elastic differential
cross section \cite{bou2014a}.

The paper is organized a follow:
in section 2, the role of the thermodynamical potentials is discussed, 
in section 3
we consider the entropy of quarks states and  show that the parameters values
obtained from a fit correspond to a maximum entropy of these states. The same
property is derived for the structure functions $F^2_p$ and $g^1_p$, and also
for the quarks.
In section 4 an analysis of the polarized gluon leads
to define a neural model for its structure and in section 5 we developp an 
extension of this model to quarks.
\section{The role of the thermodynamical potentials}
In the statistical model the thermodynamical helicity dependent
potentials $X^{\pm}$ play an essential
 role in the construction of the polarized quark distributions and so have a 
direct consequence on the behavior of the polarized structure functions. 
The helicity decomposition of the quarks  PDF is given by a quasi Fermi-Dirac 
distribution which is defined
at the input scale $Q^2 = 1$GeV$^2$ by the expressions:
\begin{equation}
xq^{\pm}(x) = \frac{A_q X_{q}^{\pm}x^{b_q}} {\exp[ (x - X_{q}^{\pm})/ 
\bar x] + 1 }
+\frac{\tilde{A}_{q}x^{\tilde{b}_{q}}}{\exp(x/\bar{x})+1}\,,
\label{quark}
\end{equation}
\begin{equation}
x\bar{q}^{\pm}(x) = \frac{\bar{A}_q}{X_{q}^{\mp}}\cdot\frac{x^{\bar{b}_q}}{ 
\exp[ (x + X_{q}^{\mp})/ \bar x] + 1 }
+\frac{\tilde{A}_{q}x^{\tilde{b}_{q}}}{\exp(x/\bar{x})+1}\,.
\label{qbar}
\end{equation}
The last term is a diffractive contribution whose effect is to enhance the
values of the unpolarized quarks at low $x$.
The polarized quarks are defined by the difference 
$x\Delta q(x) = xq^+(x) - xq^-(x)$ and
the unpolarized one by the sum $xq(x) = xq^+(x) + xq^-(x)$,
the antiquarks $\bar q$ have a similar definition. 
In these expressions $\bar x$ is a universal temperature, 
the introduction of the potential $X^{\pm}$
and $(X^{\pm})^{-1}$ in front of Eqs. (\ref{quark}-\ref{qbar})
comes from the relation
between the Transverse Momentum Distribution and the PDF \cite{tmd}.
In a recent fit of unpolarized and polarised data
made in Ref.  \cite{bou2015b} we have obtained for the $u, d, s$ 
potentials,
\footnote{The PDF are evolved with the HOPPET program \cite{hoppet}.
See Ref. \cite{bou2015b} for more details.
A Fortran program to compute the PDF independently of HOPPET
is available upon request.}
\begin{eqnarray}
&&X_u^+ = 0.475 \pm 0.0008 ,\quad X_u^- = 0.307 \pm 0.001,
\quad X_d^+ = 0.245 \pm 0.001,  \nonumber \\
&&X_d^- = 0.309 \pm 0.001,\quad X_s^+ = 0.011 \pm 0.0008,
\quad  X_s^- = 0.015  \pm 0.001\,.
\label{potval}
\end{eqnarray}
From these values we obtain the potentials hierarchy
\begin{equation}
 X_s^+ < X_s^- < X_d^+ < X_u^- < X_d^- < X_u^+ \,,
\label{inequali}
\end{equation}
which is responsible of the quarks order of magnitude.
Notice that the $u$ and the $d$ potentials  are relatively stable since 
the analysis made in 2002 \cite{bou2002},
it means that their values are a real intrinsic property of quarks,
and they represent the master parameters of the statistical model.
\begin{figure}[htp]
\vspace*{-10mm}
\begin{center}
  \epsfig{figure=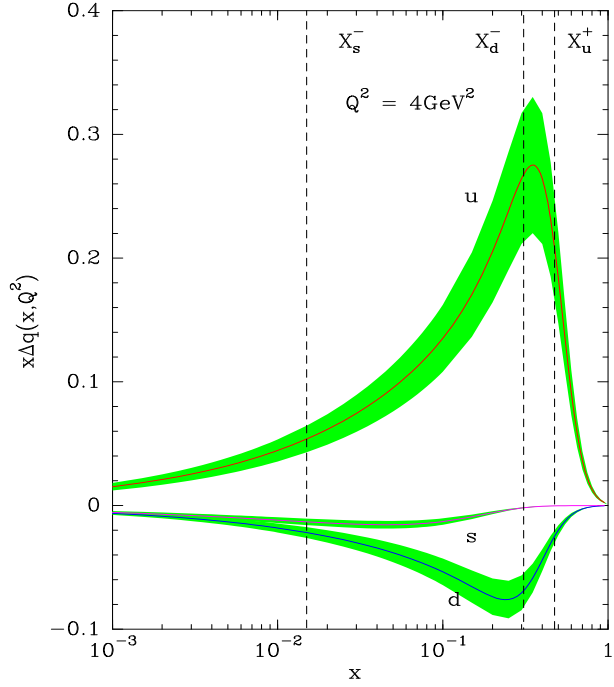,height=9cm,width=8cm}
\caption{(color online) The polarized PDF at $Q^2 =4\mbox{GeV}^2$
as a function of $x$. Vertical lines represent the maximum
value of the potential for each quark. Shaded area uncertainty bands.}
\label{polq}
\end{center}
\end{figure}
In Fig. \ref{polq} a plot of the polarized light quarks at $Q^2 = 4\mbox{GeV}^2$
is shown with their corresponding maximum potentials, we observe a correlation 
between
the potentials values and the maximum or the minimum of the PDF.
From Eq. (\ref{quark}) the sign of the polarized PDF $\Delta q = q^+ - q^-$
is related to the value of the thermodynamical
potential helicity, more precisely, on the relative
values of the potentials $X^+$ and $X^-$, the equality being excluded
because it leads to a vanishing polarized PDF at the input scale, so we are led 
with two possibilities $X^- >  X^+$ and $X^- <  X^+$.
In the case of $u$ quark $X_u^- < X_u^+$
so $\Delta u$ is positive, while for the $d$ quark $X_d^- >  X_d^+$ so 
$\Delta d$ is negative, and for the strange quark $s$ we have $X_s^- >  X_s^+$ 
leading also to a negative $\Delta s$ (see HERMES experiment \cite{hermes}). 

For the antiquarks, the chiral structure of QCD gives an important 
property which allow to relate quark and antiquark distributions. 
The potential of an  antiquark $\bar q_i^{~-h}$ of helicity {\it -h}
is opposite to the potential of a quark $q_i^{h}$ of helicity {\it h}
\begin{equation}
X_{0\bar q}^{-h} = -X_{0q}^h\,.
\label{3}
\end{equation}
So in the expression given by Eq. (\ref{qbar}) the thermodynamical
potentials have been interchanged with respect to 
the helicity $\pm$ and their sign taken to be opposite. 
This change of sign is due to the fact that a $q~\bar q$ pair
can be created by a gluon through the process  $g \rightarrow q + \bar q $. 
Due to the interchange of the potentials the sign of
$\Delta \bar u$  is positive while $\Delta \bar d$, $\Delta \bar s$ keep their
negative sign. The respective signs are confirmed by  the
parity violating asymmetry $A_L^{PV}$ measured by the STAR polarized experiment 
\cite{star} in the process $ \overrightarrow{p} p\rightarrow W^{\pm} +X$.

\begin{figure}[htb]
\begin{center}
  \epsfig{figure=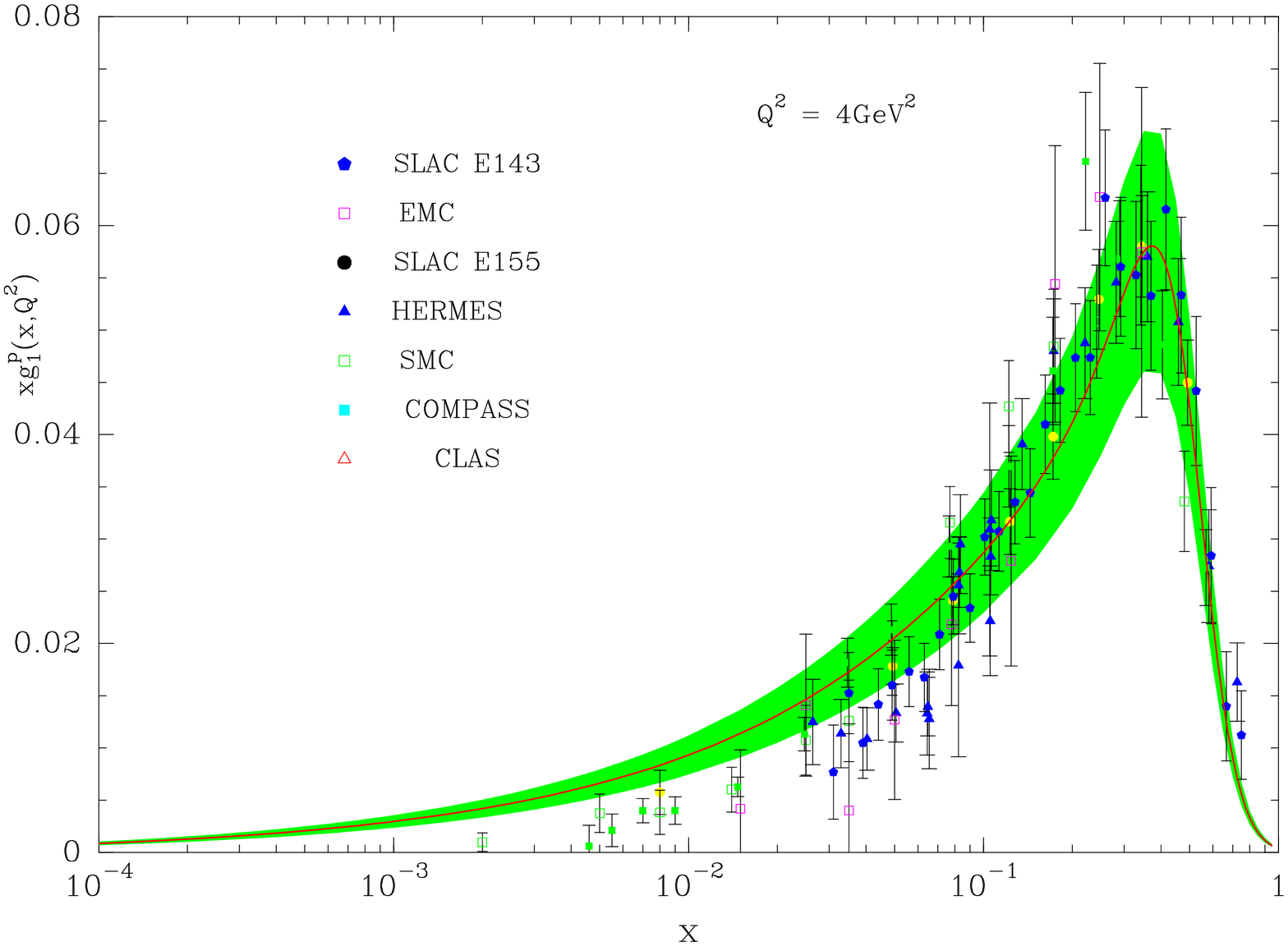,width=10.0cm}
\end{center}
  \vspace*{-5mm}
\label{xg1p}
\vspace*{-1.0ex}
\begin{center}              %
  \epsfig{figure=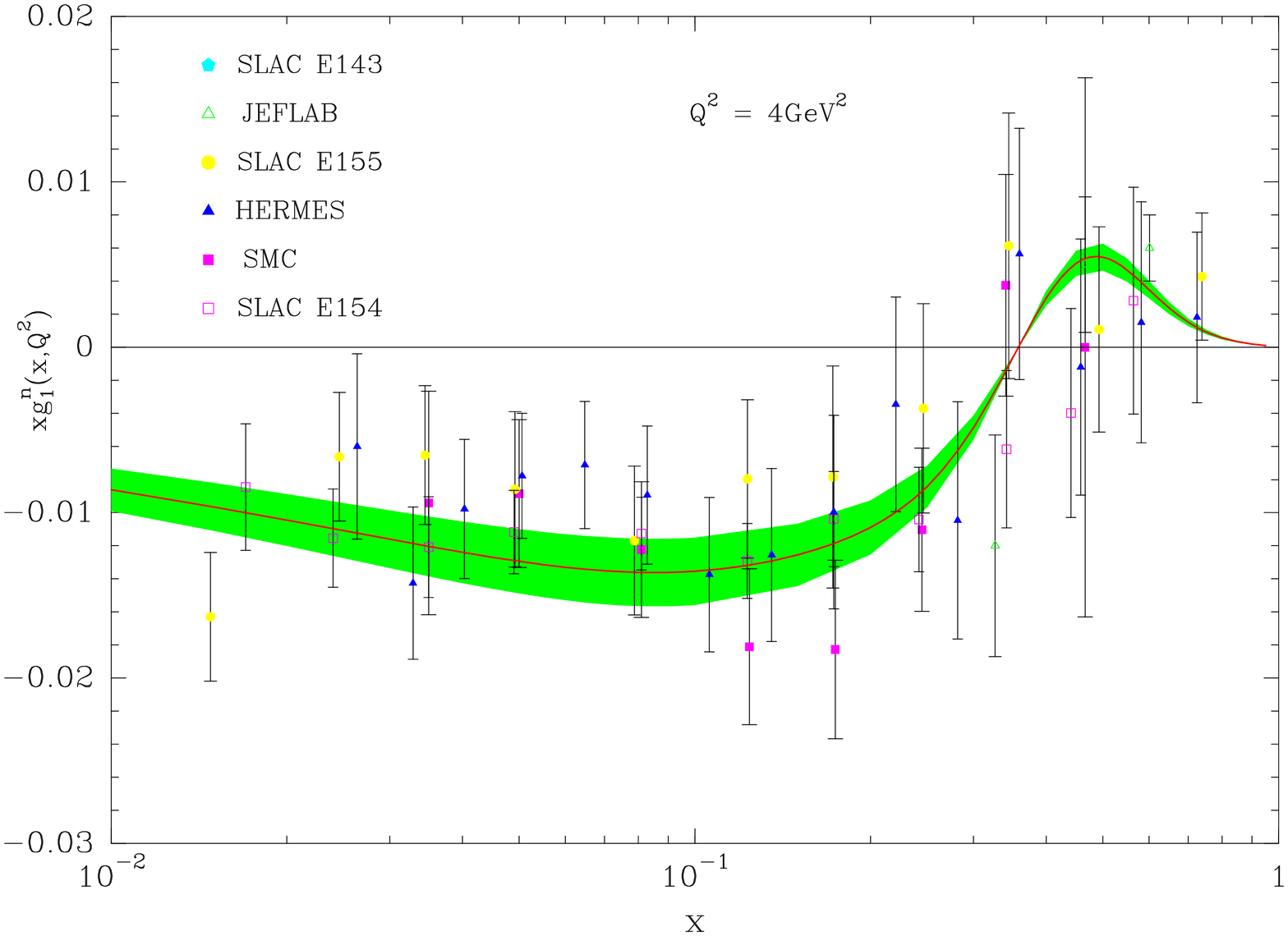,width=10.0cm}
\end{center}
  \vspace*{-5ex}
\caption[*]{(color online) The polarized structure functions $xg_1^p$ (top)
and $xg_1^n$ (bottom) as a function of $x$ for $Q^2 =4\mbox{GeV}^2$. 
Experiments \cite{emc88a}-\cite{comp10a}. Shaded area uncertainty bands.}
\label{xg1n}
\vspace*{-1.0ex}
\end{figure}

Taking into account the numerical value of the potentials how they influence
the spin structure functions, we  will give two exemples.
The polarized structure function $xg_1^p$ has a maximum around $x = 0.4$,
see Fig. \ref{xg1n}, now we know that $\Delta u$ which gives the 
major contribution has a thermodynamical potential $X_{0u}^{+} = 0.46$, 
so  we observe a correspondance between this potential and the maximum in 
$xg_1^p$.
An other exemple is given by $xg_1^n$, 
the data show that  $g_1^n$ is mainly negative over a large $x$ region this
fact can be explained by the inequality of the thermodynamical potentials
$ X_{0d}^{+} < X_{0d}^{-}$ which implies $\Delta( d^+ -  d^-) < 0$,
we also see in the figure that around $x = 0.45$ it has
a postive maximum which reflects the influence of the $u^+$ contribution 
at large $x$ compared to $\Delta d$ which is depressed in this region.

\clearpage
\newpage
\begin{figure}[htp]
\vspace*{-5ex}
\begin{center}
  \epsfig{figure=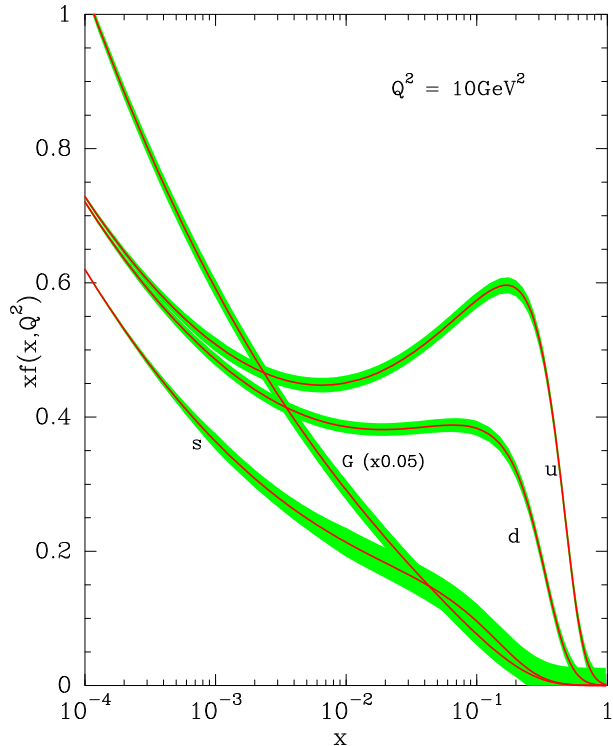,width=8.0cm}
\caption{(color online) The unpolarized PDF at $Q^2 =10\mbox{GeV}^2$
as a function of $x$. Shaded area uncertainty bands.}
\label{unpolq}
\end{center}
\end{figure}
\section{The quarks entropy}
In the previous section we have focused on the polarized PDF, 
the unpolarized ones are obtained from the relation $q = q^+ + q^-$,
we show in Fig. \ref{unpolq} the unpolarized PDF at $Q^2 = 10\mbox{GeV}^2$.
With the parameters defined above
they  give a good description of the unpolarized structure functions
in deep inelastic scattering, the neutrino cross sections, the neutral and 
charged current cross sections, and the jets production up to LHC energy,
see Ref. \cite{bou2015b}.

\begin{figure}[htp]
\begin{center}
  \epsfig{figure=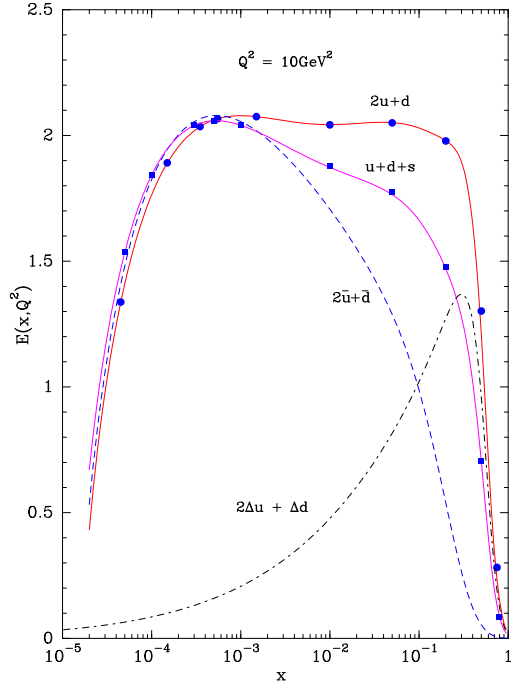,height=9.0cm}
\end{center}
  \vspace*{-5mm}
\caption{(color online) Entropy at $Q^2 = 10\mbox{GeV}^2$
as a function of $x$ for the states $|2u +d>$, $|u +d +s>$,
$|2\bar u +\bar d>$ and $|2\Delta u + \Delta d>$, 
calculated with the experimental parameters.
The optimal solutions of the entropy correspond to circles for $|2u+d>$,
and squares for $|u +d +s>$.} 
\label{entropp73}
\vspace*{-1.0ex}
\end{figure}

We will explore a new property  of the unpolarized PDF
by considering a physical quantity precisely the entropy.
The entropy can be calculated according to the definition given in
Ref. \cite{cleym12} Eq. (16)
\begin{equation}
E(Q^2,x) = -\sum_i \left[x q_i(Q^2,x) \ln{(xq_i(Q^2,x))} +(1- xq_i(Q^2,x))
\ln{(1- xq_i(Q^2,x))}\right]\,,
\label{entropy}
\end{equation}
where the sum runs over the quark components. 
We first remark that the vanishing
of $xq_i(Q^2,x)$ for $x = 1$,  implies  that $E(Q^2,x) = 0$ in this limit.
We propose to compute the entropy for the states made 
with  $|2u +d>$, $|u +d +s>$ and $|2\bar u +\bar d>$, at a fixed 
$Q^2 = 10\mbox{GeV}^2$ as a function of $x$.
In Fig. \ref{entropp73} we see that the
first state is largely dominant over the last ones
which seems to reflect the importance of matter over anti-matter.

The curves shown in  Figure  \ref{entropp73}  have been calculated with
the values of the potentials obtained from a fit of experimental data discussed 
above, then a question arises, 
what is the origin of these values, does exist a possibility to obtain them
independently of experimental data? In the PDF formulas described above we have
introduced the following  parameters: a normalization $A$, 
a power $b$ of the variable $x$,
a temperature $\bar x$ and the potentials. For a matter of simplification 
in the calculation let us assume 
that the following parameters $A,~b,~\bar x$ are held fixed to their actual 
values and consider now the potentials as free parameters
which will be determined from a calculation
of the optimal value of the entropy (\ref{entropy}) 
for a given value of $x$ and $Q^2$.
In complete generality all the the parameters of the model should have been 
considered as free, but due the complexity of the computation we restrict our
search only to the six potentials defined as the master parameters of the model.

For this purpose we consider $E(Q^2,x)$ given by Eq. (\ref{entropy}) as an 
{\it objective function} which depends on $u$, $d$, $s$ quarks subjects to the
constraints
\begin{eqnarray}
&& 0 < X_{0q}^h < 1,\quad\quad \int u_v(x) dx = 2, \quad\quad \int d_v(x) dx = 1, 
\no \\
&& \Delta u(x) \ge 0, \quad\quad \Delta d(x) \le 0,  \quad\quad \Delta s(x) \le 0,
  \no \\
&& \int (xu(x) +xd(x)+ xs(x))dx \le 1, 
\quad\quad \int (s(x) -\bar s(x))dx =0 \,.
\label{constrain}
\end{eqnarray}

The goal is to solve the sytem of equations 
(\ref{entropy})-(\ref{constrain}) with respect to the thermodynalical
potentials (supposed to be unknown) associated with $u$, $d$ and $s$.
The optimization is performed with the NLOPT software \cite{nlopt}, which involves
the objective function, the constraints and their gradients with respect
to the parameters.
In addition, to confirm the results a brute-force method is also applied, it
consists to find the maximum of the entropy by varying the parameters 
in a range defined as $\pm 50\%$ of the fitted parameters values. 

We consider for a fixed $Q^2 = 10\mbox{GeV}^2$
a set of 20 $x$ values in the range
$10^{-3} < x < 1$, the solutions  for the optimal entropy are
shown in  Fig. \ref{entropp73} as circles for the state $|2u+d>$,
and squares for the state $|u+d+s>$, one observes that their values
are close to the solids curves. 
These results show that the parameters obtained by this method have the
same values (with an error around 2\%)
as the original ones obtained from a fit, so the entropy obtained from
experimental data satisfies an optimal principle.

We can envisage also to compute the entropy of a polarized state 
$|2\Delta u +\Delta d>$, in this case
there is a the difficulty which
comes from the fact that to polarize, for instance, a proton one needs to apply
a strong magnetic field, so there is coupling
between the proton and an external field which introduces a complicated situation
for the computation of the entropy because one has to disantangle the 
contribution coming from the external field and the other 
from the state itself.
\begin{figure}[htp]
\begin{center}
  \epsfig{figure=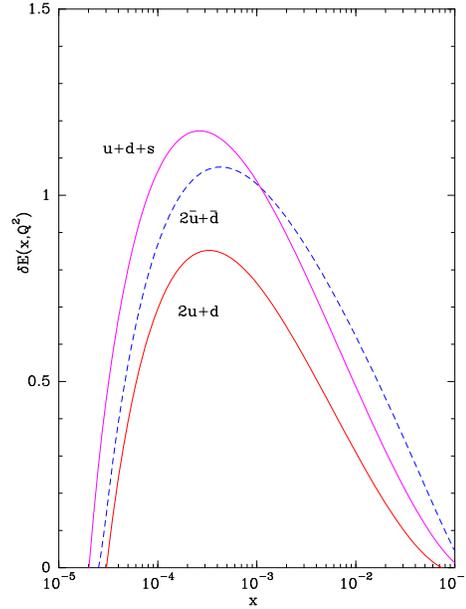,height=8.0cm}
\end{center}
  \vspace*{-5mm}
\caption{(color online) Difference of the entropy  at
$Q^2 = 1 ~\mbox{and}~ 10\mbox{Gev}^2$.}
\label{diffentrop}
\vspace*{-1.0ex}
\end{figure}
 Nevertheless, we show in Fig. \ref{entropp73} the resulting
entropy with a dash-dotted curve, the values are much smaller than in
the case of an unpolarized proton where the situation is more clear
because the proton can be considered as in a free state.
As a conclusion
the calculation of the entropy obtained in an independent way from experiment has the
consequence that the quarks PDF obtained from a fit correspond to
a maximum entropy principle, so the structure functions must share 
in some way the same property.
We know that the entropy is sometimes associated with the disorder of a system 
and increases with energy. 
In Fig. \ref{diffentrop} the difference of the entropy  $\delta E$ between
$Q^2 = 1 ~\mbox{and}~ 10\mbox{GeV}^2$ is effectively growing for the
states discussed above. The state which involves the strange quark has the 
largest effect with respect to the disorder.

A comparison of the optimum calculated entropy with experiment is not easy,
an other test can be made with the structure functions. Using the same method
as above, we consider again the thermodynamical potentials as free parameters
in a certain range of values and search a maximum for the strcture functions
$F_2^p$ and $g_1^p$ for a fixed $x$ and $Q^2$. We find that the maximum
is obtained when the potentials values are those obtained in the fit, so the
following relation is derived
\begin{equation}
 F_2^p~\mbox{maximum} =  F_2^p~\mbox{fit} =  F_2^p~\mbox{experimental}\,,
\label{optimum}
\end{equation}
with the same relation for  $g_1^p$. 

\begin{figure}[tbp]  
  \begin{minipage}{8.5cm}
  \epsfig{figure=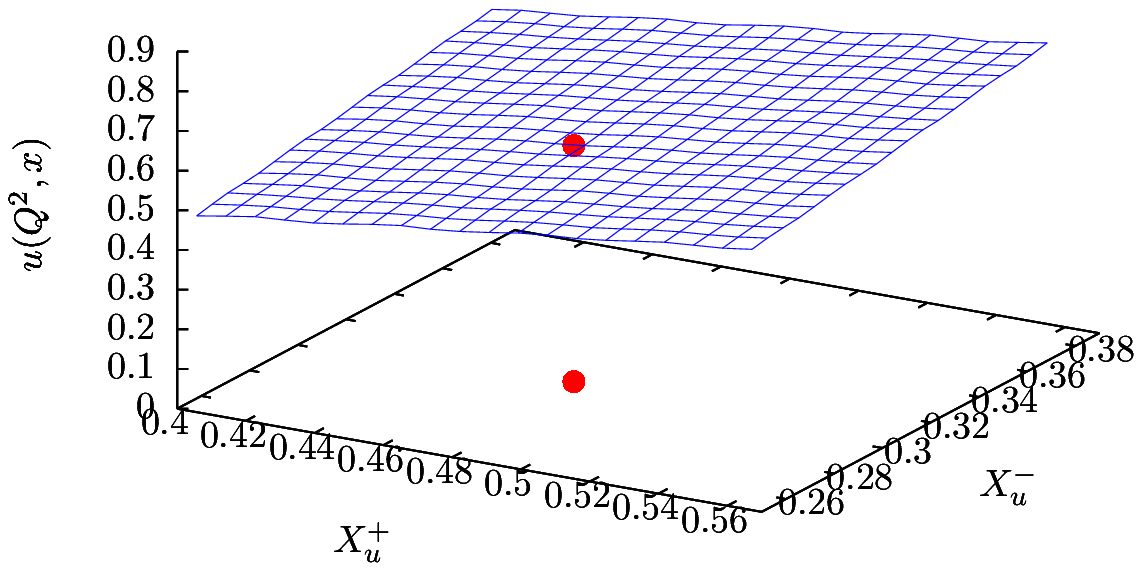,width=8.5cm}
  \vspace*{-5mm}
  \end{minipage}
  \hspace*{-15mm}
    \begin{minipage}{8.5cm}
  \epsfig{figure=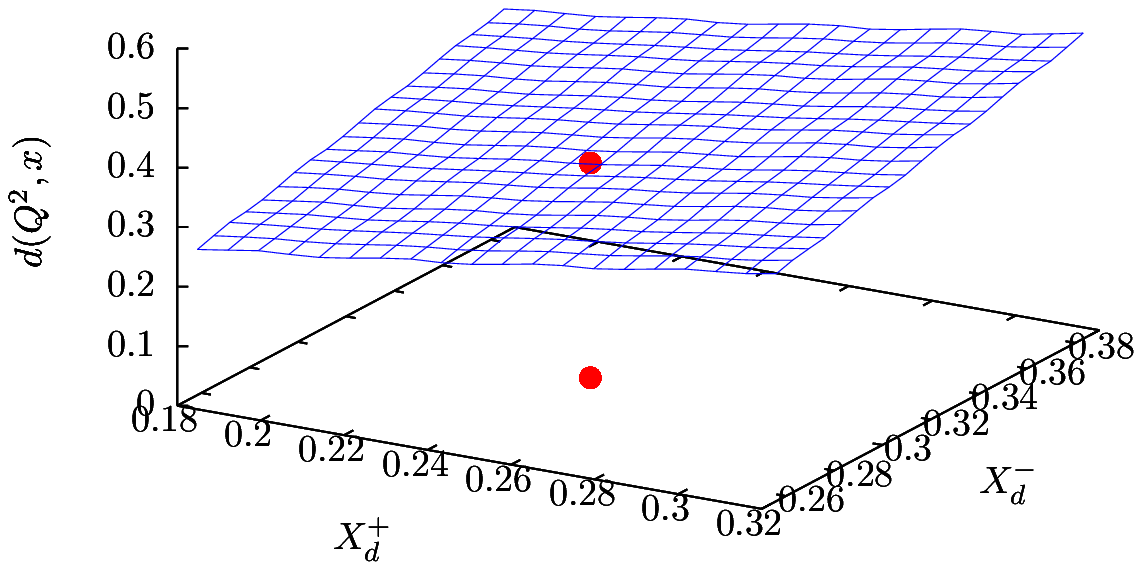,width=8.5cm}
  \vspace*{-5mm}
    \end{minipage}\\
\caption{
(color online) Quark distributions as a function of $X^+$, $X^-$ for 
$Q^2 = 10\mbox{GeV}^2$ and $x = 0.15$, {\it left} up-quark,
{\it right} down quark. Maximum value red point.}
\label{qmax}
\end{figure}

In the above results the quark distributions are the essential source
of information to obtain an optimum a property which should be reflected in the quarks
themselves. To prove this we consider that in the unpolarized up and down quarks 
the potentials are now free parameters and a search is made for a maximum value
given a fixed $Q^2 = 10\mbox{GeV}^2$ and $x = 0.15$. 
In Fig. \ref{qmax} we plot the u and d values as a function
of $X^+$ and $X^-$ limited to a certain domain which  generates a surface.
Now, if we look  for  a maximum by imposing 
the constraints 
Eqs . \ref{constrain} we obtain as a solution only one couple $X^+, X^-$ with
a $u$ and $d$ values which correspond to those obtained in the fit (red point
in the figure). The same result is also obtained for the polarized $\Delta u$ and
$\Delta d$. So the optimum obtained  for the entropy and the structure functions
find its origin on the quarks properties. Let us mention that this maximum values
of the light quarks unpolarized distriutions is also found with the 
parametrization MSTW 2008 \cite{mstw} and CT14 \footnote{I thank J. Pumplin for informations on CT14 PDFs.}
\cite{ct14}.

From this result we infer that nature
tends to produce observable quantities with a maximum probability  taking into
account some physical constraints, it remains to explain the origin of this effect.

\section{A neural model of  polarized gluon}
The polarized gluon distribution is today  not well known and subject to a large
debate concerning its expression and sign.
Our purpose is to clarify the choice made in our original model at the input scale
and to propose a new interpretation in the context of a neural structure.
In the statistical model it is natural to assume a quasi Bose-Einstein 
distribution for both $G$ and $\Delta G$, so we define for the gluon at the input
scale $Q_0^2 = 1\mbox{Gev}^2$
\begin{equation}
xG(x,Q^2_0) = \frac{A_G x^{b_G}}{\exp(x/\bar{x})-1}\,,
\label{eq3}
\end{equation}
and for the polarized distribution
\begin{equation}
 x\Delta G(x,Q^2_0) = \tilde A_G x^{\tilde b_G}  P(x)\cdot
\frac{1}{\exp(x/\bar x - 1) } \,,
\label{polglue}
 \end{equation}
where for $P(x)$ we made the choice \cite{bou2015a}
\begin{equation}
P(x) = \frac{1}{(1+ c_G x^{d_G})}\,,
\label{funcpx}
\end{equation}
notice that
the introduction of an analogous  rational multiplicative factor 
is also used in Ref. \cite{florian2014}.
A fit of polarized DIS data gives the values \cite{bou2015b}
\begin{eqnarray}
\nonumber
A_G = 36.778 \pm 0.085,~b_G = 1.020 \pm 0.0014,~\tilde {A}_G = 26.887 
\pm 0.050,\\
\tilde {b}_G = 0.163 \pm 0.005,~c_G = 0.006 \pm 0.0005, ~d_G = -6.072 
\pm 0.350\,,
\label{eq8}
\end{eqnarray}
and for the temperature $\bar x =0.090 \pm 0.002$. We obtain a 
$\chi^2/d.o.f. = 319/269$.
\begin{figure}[htp]
\vspace*{-5ex}
\begin{center}
  \epsfig{figure=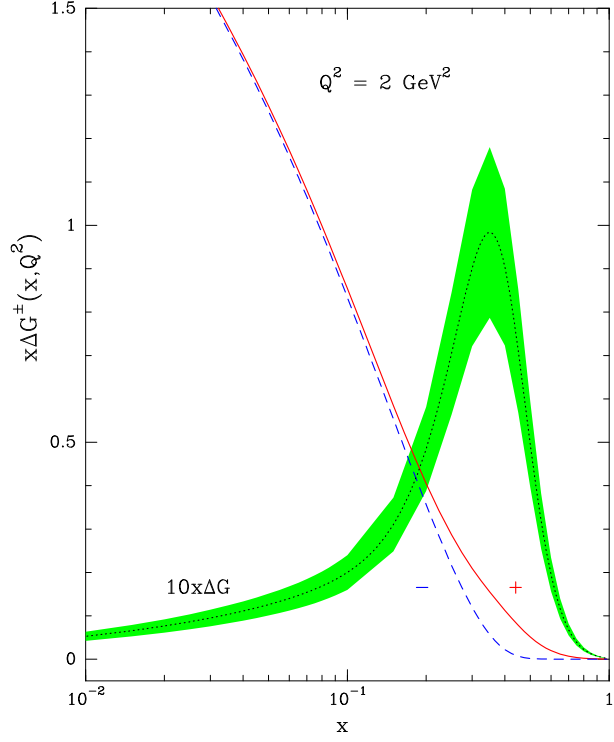,width=8.0cm}
\end{center}
  \vspace*{-8mm}
\caption{(color online) $x\Delta G(x, Q^2)$  at $Q^2 =2\mbox{GeV}^2$, dotted curve 
values multiplied by 10, 
helicity components solid + and dashed curves -, as a function of $x$.
Shaded area uncertainty bands.}
\label{gluonhel}
\vspace*{-2.0ex}
\end{figure}
\begin{figure}[htp]   
\vspace*{-4.0ex}
\begin{center}
\includegraphics[width=11.0cm]{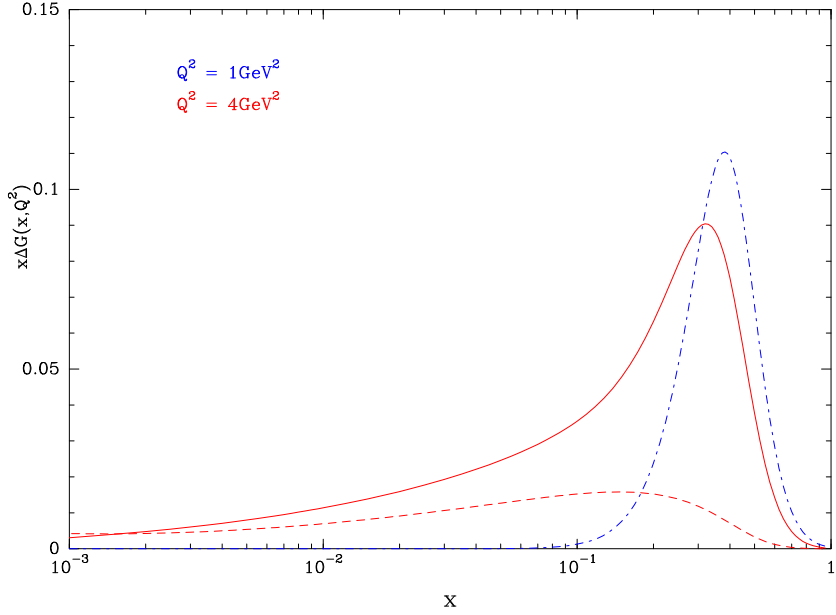}
\caption[*]{\baselineskip 1pt
 (color online) The gluon helicity distribution $x\Delta G(x,Q^2)$ versus $x$, at
$Q^2=1\mbox{GeV}^2$ (dash-dotted curve) and $Q^2=4\mbox{GeV}^2$  with
$c_G \ne 0$ (solid curve), and $c_G =0$ (dashed curve).}
\label{gluon-hel}
\end{center}
\end{figure}
\begin{figure}[htp]
\begin{center}
\includegraphics[width=11.0cm]{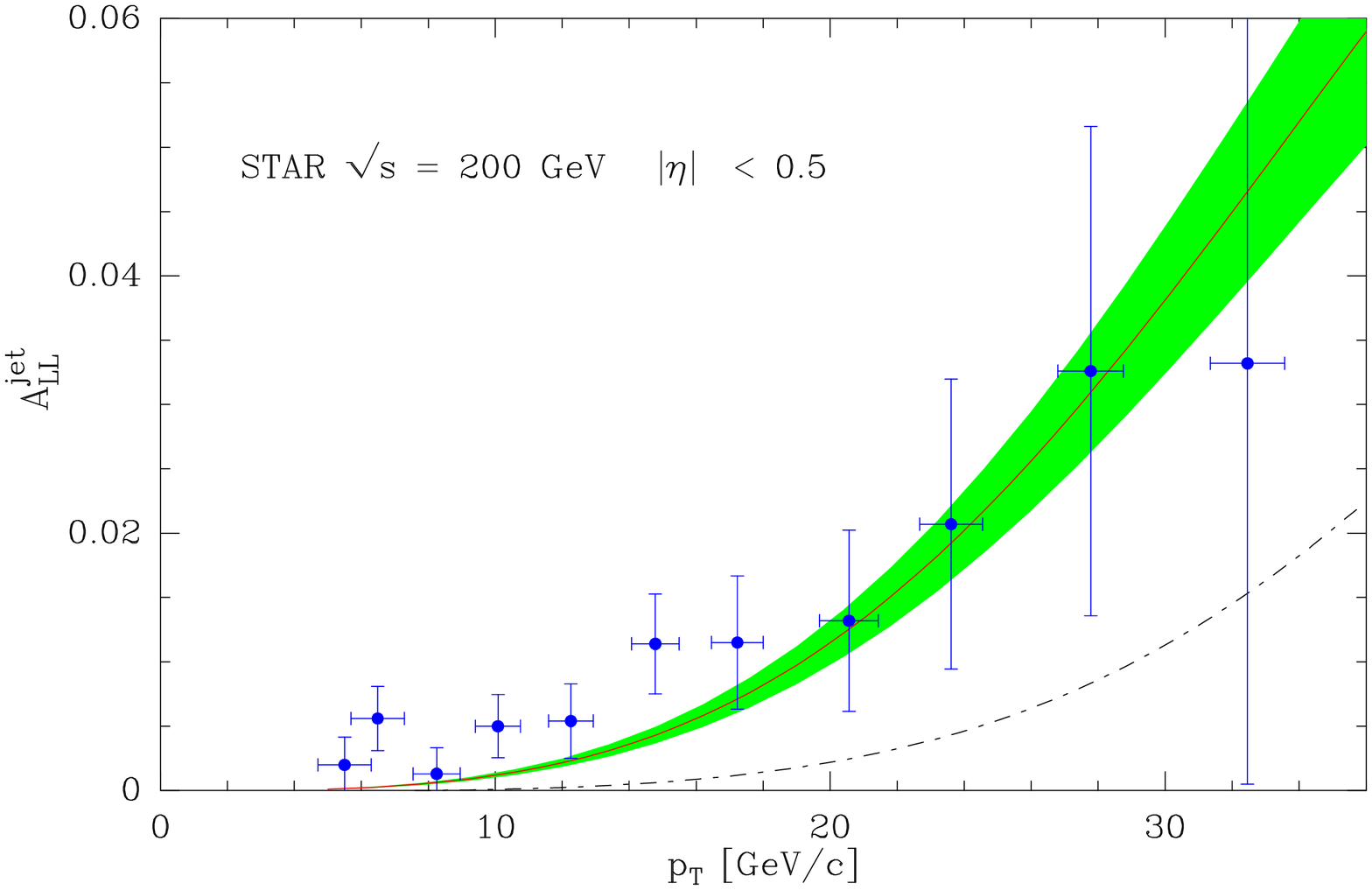}
\caption[*]{\baselineskip 1pt
(color online) {\it Solid curve}: the predicted double-helicity asymmetry 
$A_{LL}^{jet}$ 
with $c_G \ne 0$ for
jet production at BNL-RHIC in the near-forward rapidity region, versus $p_T$
with the data points from Ref. \cite{star}.\\
{\it Dot Dashed curve}: the asymmetry with  $c_G =  0$.}
\label{allcg0}
\end{center}
\end{figure}
With these parameters $\Delta G(x)$ is  positive a property  confirmed by
experimental data from Hermes \cite{hermes10},
Compass \cite{compass13} and $A^{jet}_{LL}$ from the STAR Collaboration at 
BNL-RHIC \cite{star14}.
In our analysis we take advantage that both the
unpolarized and polarized gluon are determined in the same fit. 
We remark that unpolarized quarks and gluon are related 
through
the evolution equations, in the same way the polarized quarks and gluon
are related by an other set of evolution equations, now  by construction unpolarized
and polarized quarks are related because the building blocks
are the helicity components, as a consequence all the partons are linked together.

A plot in Fig. \ref{gluonhel} of the polarized gluon (dotted curve)  at 
$Q^2 = 2$GeV$^2$ shows a maximum  in the region $x = 0.3$, now from the 
values of $G(x)$ and $\Delta G(x)$
we can deduce  the helicity components (solid and dashed curves)
their main difference is located in the same $x$ region.

A priori, it was natural to use for the polarized gluon an analogous expression
like Eq. (\ref{eq3}) for the gluon which is obtained  by setting $c_G = 0$ 
in $P(x)$.
With this assumption a new fit gives for the parameters
\begin{equation}
\tilde {A}_G = (2.46 \pm 0.1) 10^{-4}, ~ \tilde {b}_G = 0.294
\label{eq8a}
\end{equation}
with a $\chi^2/d.o.f.= 325/269$, at this level we can consider that the 2 solutions
(\ref{eq8})-(\ref{eq8a})
are equivalent, however, in Fig. \ref{gluon-hel} we observe a marked difference
for $x\Delta G$. The pic obtained at $Q^2 = 4\mbox{GeV}^2$ when  $c_G = 0.006$ 
(solid curve) becomes 
a flat maximum when $c_G = 0$ and is reduced by a factor 4 (dashed curve).
In order to separate the 2 solutions we refer to the measurement of the 
double-helicity asymmetry $A_{LL}^{\it jet}$ for $5 \leq p_T \leq 30 \mbox{GeV}$, 
in the near-forward rapidity region  measured recently by the STAR 
Collaboration \cite{star}.

In Fig. \ref{allcg0} we have plotted the solution with $c_G \ne 0$ 
(solid curve)\footnote{Results are taken from Ref. \cite{bou2015a}.}
and the solution  $c_G =0$ (dashed curve). Despite large experimental errors
the agreement with data is clearly in favor
of $c_G \ne 0$ because the other curve grows too slowly with $p_T$, so the 
solution
with $c_G = 0$ is not acceptable.

\begin{figure}[htbp]   
\begin{center}
\includegraphics[width=8.0cm]{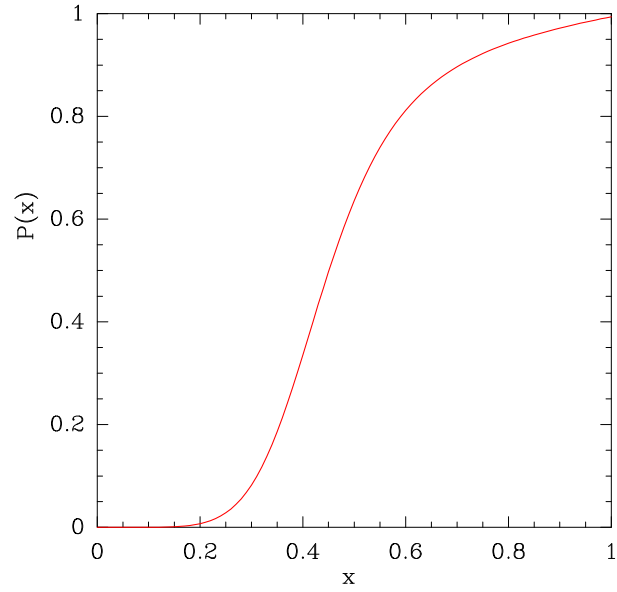}
\caption{(color online) The function $P(x)$ (Eq.(\ref{funcpx})) versus $x$
for the polarized gluon.}
\label{plotpx}
\end{center}
\end{figure}

Let us now examine the function $P(x)$ defined by Eq. (\ref{funcpx}). 
In Fig \ref{plotpx} a plot  versus $x$ at the input scale
shows that $P(x)$ is increasing with $x$, its first derivative 
is maximum for $x = 0.41$ and the second derivative (curvature) vanishes at the 
same $x$ value.
The shape of curve and the above properties show a close similarity with 
a sigmo{\"i}d  or logistic function whose basic expression is
\begin{equation}
S_o(x) = \frac{1}{1 +e^{-\lambda x}}\,.
\label{sigm0}
\end{equation}
A sigmo{\"i}d function is used as an activation function in several domains, 
in particular, in neural networks applied to structure functions
\cite{nnpdf14, forte}, also in the exploration of opacity in elastic hadron
scattering \cite{menon15}.
From our previous remark will consider that the $P(x)$ function can now be 
replaced by a sigmo{\"i}d of the form
\begin{equation}
S(x) = \frac{1}{1 +e^{-e_G x + h_G}}\,,
\label{sigmv}
\end{equation}
where the parameter $h_G$ defines a translation of the curve in the interval 
$[0\leq x  \leq 1]$, so we define a new $x\Delta G$ at the input scale
\begin{equation}
 x\Delta G(x,Q^2_0) =  S(x) \frac{\tilde {A'}_G x^{b_G}}{\exp(x/\bar{x})-1}\,.
\label{polglues}
 \end{equation}
A fit of polarized data yields the values
\begin{equation}
 \tilde {A'}_G = 18.987\pm 1.5, ~ e_G = 77.94 \pm 3.3, ~ h_G = 22.18 \pm 2.4\,,
\label{eq8b}
\end{equation}
$b_G$ is the same as in (\ref{eq3}),
we obtain a $\chi^2/d.o.f. = 326/271$ very close to the original solution.

\begin{figure}[htbp]   
\vspace*{-5.5ex}
\begin{center}
\includegraphics[width=8.0cm]{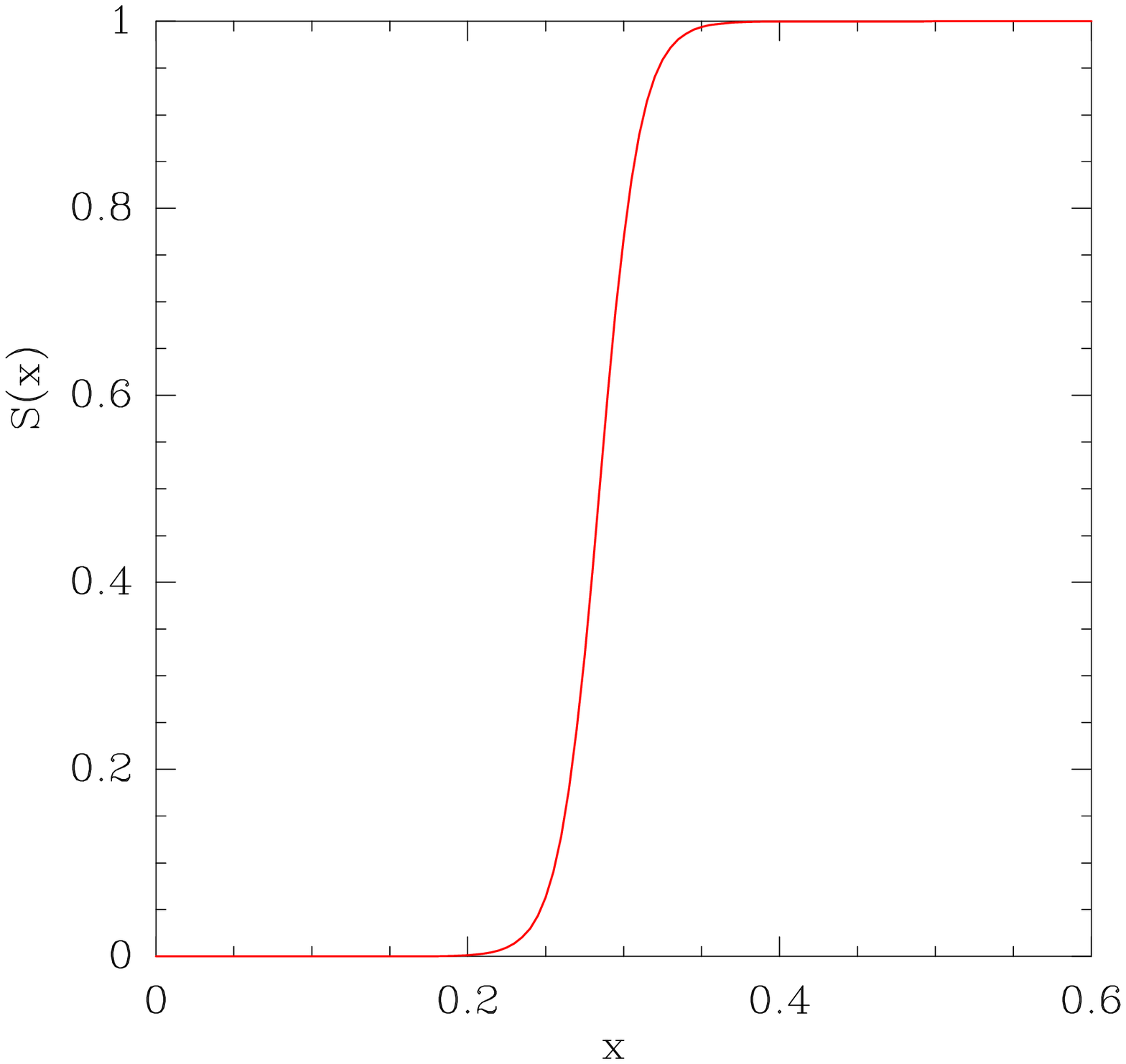}
\vspace*{-1.5ex}
\caption{(color online) The activation function $S(x)$ (Eq. (\ref{sigmv}))
versus $x$ for the polarized gluon .}
\label{plotsx}
\end{center}
\end{figure}

\begin{figure}[htpb]   
\vspace*{-5.0ex}
\begin{center}
\includegraphics[width=9.0cm]{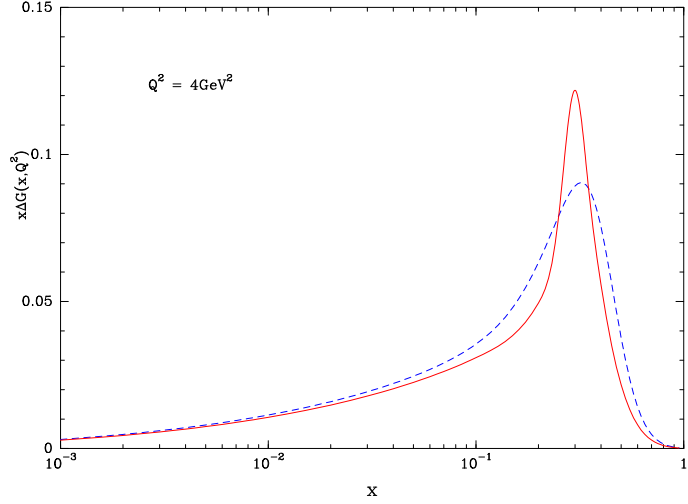}
\caption[*]{\baselineskip 1pt
(color online)  The gluon helicity distribution $x\Delta G(x,Q^2)$ versus $x$, at
$Q^2=4\mbox{GeV}^2$  calculated with $P(x)$ (dashed curve)
and with $S(x)$ (solid curve).}
\label{delgsx}
\end{center}
\end{figure}

For illustration we show in Fig. \ref{plotsx} the function $S(x)$.
calculated with the fitted parameters (\ref{eq8b}).
It is interesting to compare $x\Delta G$ obtained with $P(x)$ and $c_G \ne 0$
with the case of $S(x)$. In Fig. \ref{delgsx} the solution with  $P(x)$ 
corresponds to the dashed curve and with $S(x)$ solid curve, we observe 
that the latter has a more pronounced peak which decreases with  $Q^2$
and moves slowly toward smaller $x$ values.

\begin{figure}[hpb]   
\vspace*{-5.0ex}
\begin{center}
\includegraphics[width=11.0cm]{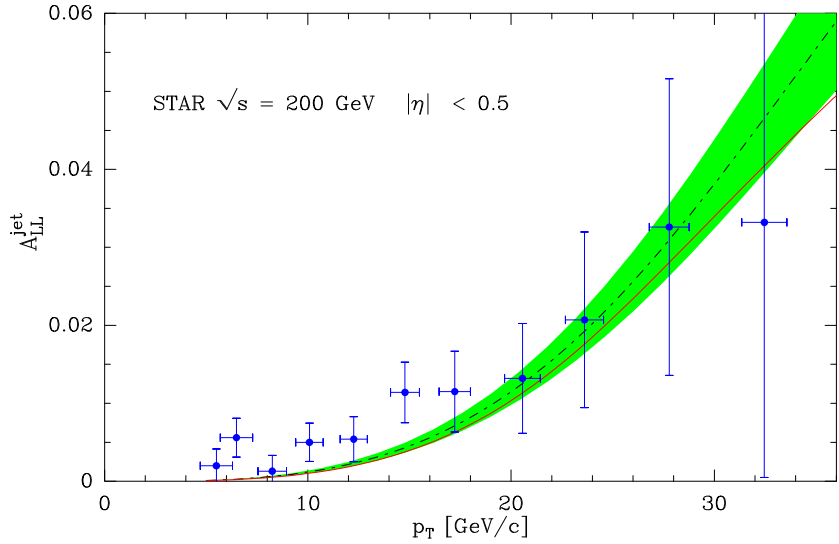}
\caption[*]{\baselineskip 1pt
(color online) {\it Dot dashed curve}: Our predicted double-helicity asymmetry 
$A_{LL}^{jet}$ with $P(x),~c_G \ne 0$ for
jet production at BNL-RHIC in the near-forward rapidity region, versus $p_T$
and the data points from Ref. \cite{star}.\\
{\it  Solid curve}: the asymmetry calculated with the function $S(x)$.}
\label{allsigmo}
\end{center}
\end{figure}

The validity of the new polarized gluon can be tested by computing the
asymmetry $A_{LL}^{jet}$, the Fig. \ref{allsigmo} shows a good agreement
between the two solutions, $P(x), c_g \ne 0$ and $S(x)$.

In this section we have explored three possibilities to describe the helicity
of the gluon. Starting with the expression  Eq. (\ref{funcpx}) used in 
\cite{bou2015b}
which was phenomelogical, now we have shown that a more physical expression
given by a sigmo{\"i}d Eq. (\ref{sigmv}) gives also a good desciption of polarized
experimental data. 

\begin{figure}[hbp]   
\begin{center}
\includegraphics[width=8.0cm]{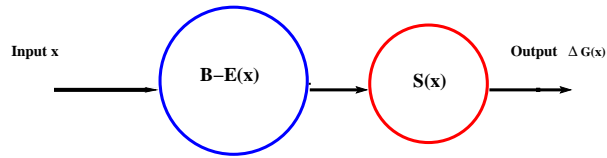}
\caption[*]{\baselineskip 1pt (color online)
A schematic representation of a polarized gluon in a neural model.
$B-E(x)$ is the Bose-Einstein distribution, $S(x)$ the activation function.}
\label{neural}
\end{center}
\end{figure}

\begin{figure}[htbp]
\begin{center}
\includegraphics[width=8.0cm]{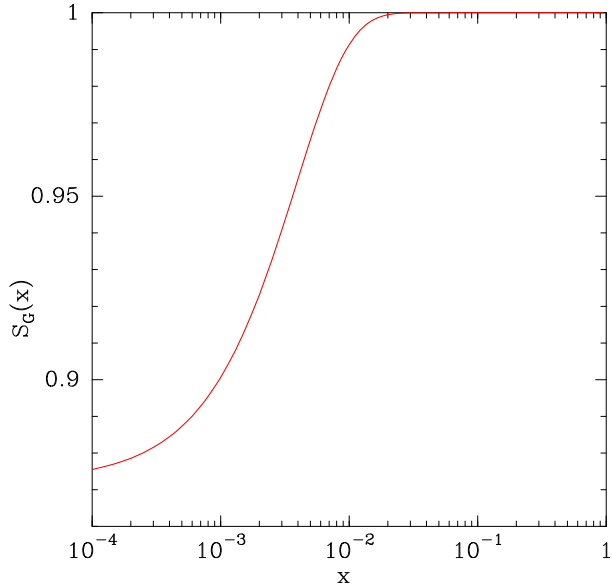}
\vspace*{-1.5ex}
\caption{(color online) The activation function
$S(x)$ versus $x$ for the unpolarized gluon.}
\label{sigmoglu}
\end{center}
\end{figure}

How we can interpret the role of the sigmo{\"i}d $S(x)$.
When 2 protons collide the gluon receives different fractions of the momentum
coming from the quarks which are collected
statistically with a Bose-Einstein distribution, next the function $S(x)$
plays the role of an activation function which synthesizes in an output signal 
$\Delta G$. 
This mechanism allow us to define a neural representation of the polarized
gluon whose schematic view is given in Fig. \ref{neural}.
From this result it is tempting to apply the same representation to the unpolarized
gluon, we now introduce in the gluon distribution an activation function (\ref{sigmv}).
A global fit gives the parameters
\begin{equation}
A_G = 33.23 \pm 1.2,~e_G = 281.16 \pm 6.1,~h_G = -1.923 \pm 0.01 \,.
\label{sigmogluon}
\end{equation}
The resulting $S(x)$ function is plotted in Fig. \ref{sigmoglu}, we observe
for $x > 10^{-3}$ a value of $S(x)$ around 0.9-1 almost independent of $x$.
It implies that the activation function makes no modification on the output distribution G,
which can be interpreted by the fact that in order to maintain the confinement of
quarks any momentum transfer is allowed, also as stated in the introduction the creation
of a maximum of $q \bar q$ pairs with increasing energy 
implies no selection.

\begin{figure}[htp]
\vspace*{-9ex}
\begin{center}
\includegraphics[width=7.5cm]{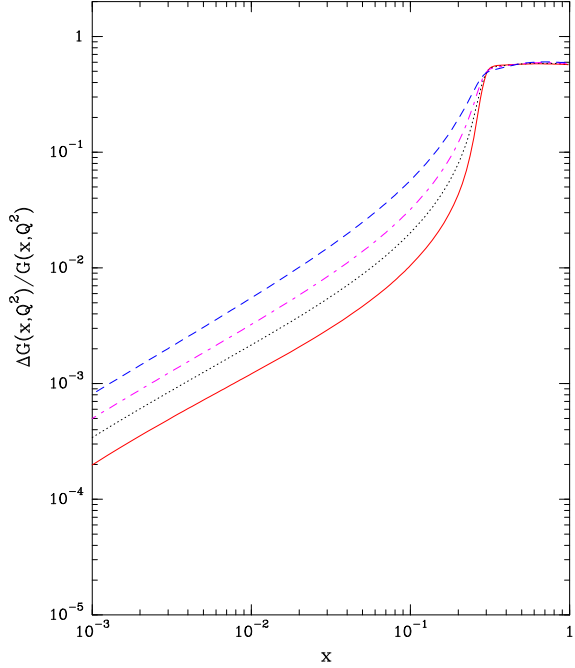}
\vspace*{-1.5ex}
\caption{(color online) $\Delta G(x, Q^2)/G(x, Q^2)$  at 
$Q^2 = 2, 4, 10, 100 \mbox{GeV}^2$ (solid, dotted, dash-dotted, and dashed 
curves respectively) as a function of $x$.}
\label{delgog}
\end{center}
\end{figure}

In this new approach of the polarized gluon we would like to examine the
ratio $\Delta G(x, Q^2)/G(x, Q^2)$ discussed in Ref. \cite{bou2015a}.
In Fig. \ref{delgog} the ratio 
is plotted as a function of $x$ for four $Q^2$ values, in this range
the positivity condition $\Delta G(x)/G(x) \leq 1$ is statisfied, and for a 
fixed $Q^2$ it increases with $x$,
near the limit $x = 1$ the values are very close to 0.5.
At the input scale
the Bose-Einstein function cancels in Eqs.(\ref{eq3})-(\ref{polglues}) it
results that
$\Delta G/G(x = 1) \rightarrow A'_G S_G(1)/A_G S_{\Delta G}(1)$, 
now taking into account the ratio of the normalization factors we obtain
 the value 0.57, so the limit of the ratio $\Delta G/G$ is different 
from 1 as required by the counting sum rule.

\section{A neural model applied to quarks}
\begin{figure}[htbp]
\vspace*{1.5ex}
\begin{center}
\includegraphics[width=8.0cm]{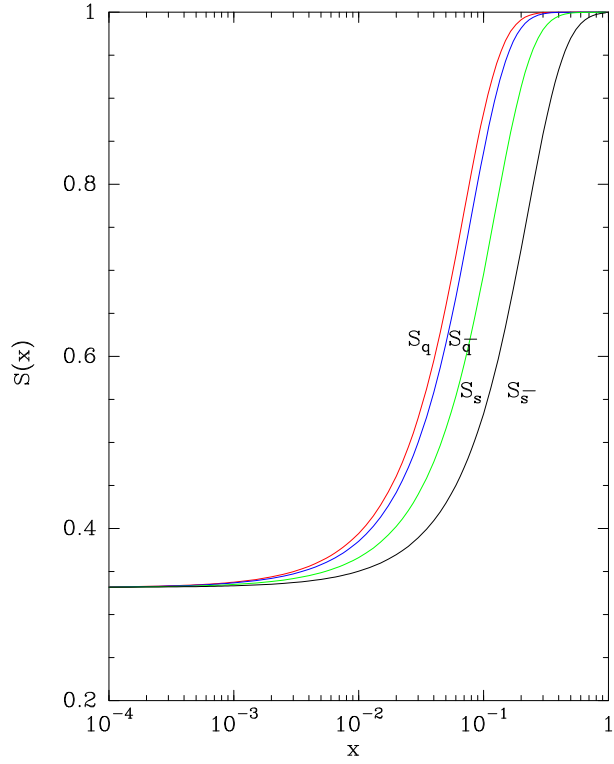}
\vspace*{-1.5ex}
\caption{(color online) The activation function
$S(x)$ versus $x$ for $q$, $\bar q$, $s$, $\bar s$,
in the domain $x \in [10^{-4},~1]$.}
\label{sigmoq}
\end{center}
\end{figure}

In the previous section
we have focused on the structure of the gluon in a neural model, 
now a question arises
for the quarks, can they share the same structure? The unpolarized quarks PDF 
are known with a good precision, and most of the parametrizations 
agree to produce the same values in $x$ and $Q^2$, in the polarized case
there are more uncertainties but the observed shapes are more or less
indentical, so the neural description  we give in the gluon case seems not
necessary. Nevertheless, looking at our PDF expressions  Eqs. (\ref{quark},
\ref{qbar}) 
we have the product of a Fermi-Dirac distribution by an helicity dependent 
funcion 
$A x^B X^{\pm}$ for quarks and $\bar A x^{\bar B}/ X^{\pm}$ 
for antiquarks, so we can try to apply the same approach  where
the incoming momentum is collected now by mean of a Ferm-Dirac distribution and then
filtered by an activation function to produce the quark distribution. 
Our objective is to obtain a coherent neural structure for all the 
unpolarized and polarized PDF.
Several possibilities exist to introduce an activation function, we
made the following choice where the original parton  expressions for 
$ q, \bar q ~\mbox{and}~ G$ are preserved when $S(x) = 1$.
\begin{equation}
xq^{\pm}(x) = S_q(x) \frac{A_q X_{q}^{\pm}x^{b_q}} {\exp[ (x - X_{q}^{\pm})/ 
\bar x] + 1 }
+\frac{\tilde{A}_{q}x^{\tilde{b}_{q}}}{\exp(x/\bar{x})+1}\,,
\label{quarkneural}
\end{equation}
\begin{equation}
x\bar{q}^{\pm}(x) = S_{\bar q}(x) \frac{\bar{A}_q}{X_{q}^{\mp}}
\cdot\frac{x^{\bar{b}_q}}{ 
\exp[ (x + X_{q}^{\mp})/ \bar x] + 1 }
+\frac{\tilde{A}_{q}x^{\tilde{b}_{q}}}{\exp(x/\bar{x})+1}\,.
\label{qbarneur}
\end{equation}
\begin{equation}
xG(x,Q^2_0) = S_G(x) \frac{A_G x^{b_G}}{\exp(x/\bar{x})-1}\,,
\label{eq3neur}
\end{equation}
\begin{equation}
 x\Delta G(x,Q^2_0) = S_{\Delta G}(x) \tilde A_G x^{\tilde b_G}  \cdot
\frac{1}{\exp(x/\bar x - 1) } \,,
\label{polgluenneur}
 \end{equation}
with an activation function defined by:
\begin{equation}
S_i(x) = \frac{1}{1 +e^{-e_i x + h_i}}\,,
\label{sigmg}
\end{equation}
the index $ i = q, \bar q, G, \Delta G$, where
$S_q$ is identical for $u$, $d$ quarks, and $S_{\bar q}$ for $\bar u,~\bar d$
To determine the parameters a global fit at NLO with an activation function included in 
all PDF is made with the same data set used in \cite{bou2015b}, we obtain a
$\chi^2/d.o.f. = 2506/2128 = 1.18$.
The thermodynamical potentials are slightly modified
\begin{eqnarray}
&&X_u^+ =0.540  \pm 0.0014 ,\quad X_u^- = 0.336 \pm 0.0012,
\quad X_d^+ = 0.268 \pm 0.0013,  \nonumber \\
&&X_d^- = 0.349 \pm 0.001,\quad X_s^+ = 0.0111 \pm 0.0011,
\quad  X_s^- = 0.0147  \pm 0.0012\,.
\label{potvalnew}
\end{eqnarray}
The activation function parameters are given in Table \ref{table1} and the 
corresponding functions $S_q$ are shown in Fig. \ref{sigmoq}.
\begin{table}[ht]
\begin{center}
\begin{tabular}{|c||c|c|}\hline
$i$ & $ e_i$& $h_i$
\\ \hline
 u, d &  27.16 $\pm$ 1.3 &  0.7 \mbox{(fixed)}
\\ \hline
$\bar u$, $\bar d$ & 23.37 $\pm$ 1.07 & ''
\\ \hline
$s$ & 15.27 $\pm$ 0.9 & ''
\\ \hline
$\bar s$ & 8.34 $\pm$ 0.5  & ''
\\ \hline
 $G$ & 281.67 $\pm$ 3.9 &  -1.82 $\pm$ 0.1
 \\ \hline
 $\Delta G$ & 77.71 $\pm$ 2.0 & 22.07 $\pm$ 1.24
 \\ \hline
 \end{tabular}
\caption{Parton parameters of the activation functions.}
\label{table1}
\end{center}
\end{table}
The curves characterize the response of partons to a signal, the momentum, we 
observe a hierarchy where the $u,~d$ quarks have the dominant effect followed 
by antiquarks,
strange and antistrange, it corresponds to the observed relative size of the PDF.
In this first approach we have used the same activation function
for $u,~d,~\Delta u,~\Delta d$, idem for the antiquarks, the strange and antistrange,
but a more refined version could introduce an activation function for each
helicity components.
\begin{figure}[htbp]
\vspace*{-5.5ex}
\begin{center}
\includegraphics[width=8.0cm]{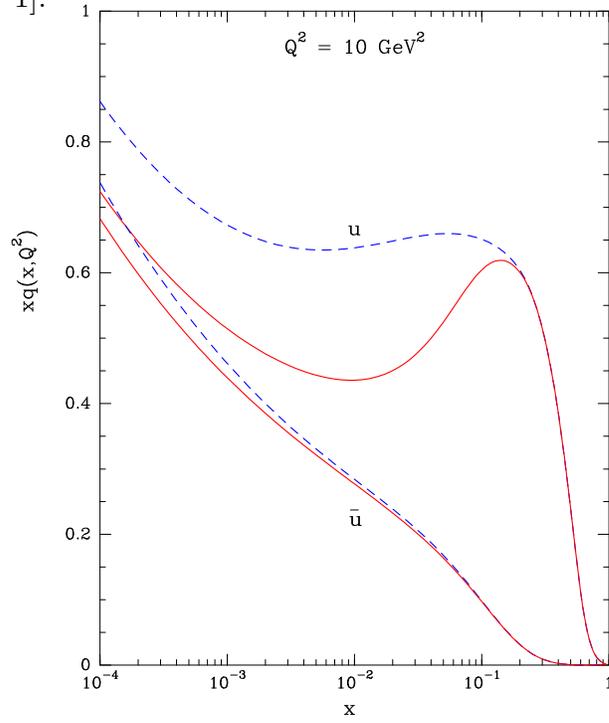}
\vspace*{-1.5ex}
\caption{(color online) Unpolarized $u$, $\bar u$ solid curves,
when $S = 1$ dashed curves.}
\label{plotq}
\end{center}
\end{figure}
\clearpage
\newpage
\begin{figure}[htbp]
\vspace*{-8.5ex}
\begin{center}
\includegraphics[width=8.0cm]{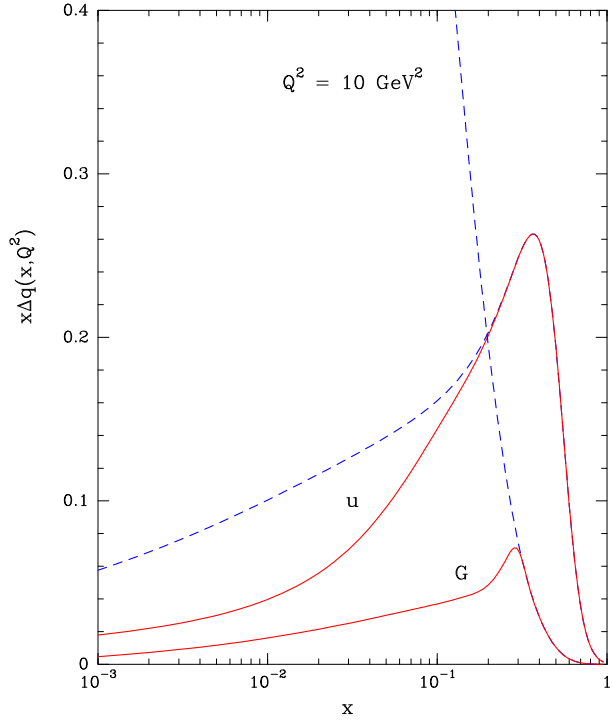}
\vspace*{-1.5ex}
\caption{(color online) Polarized $\Delta u$, $\Delta G$ solid curves,
when $S = 1$ dashed curves.}
\label{plotqpol}
\end{center}
\end{figure}
In order to show 
more precisely the  effect of the activation function we plot in
 Figs. \ref{plotq}-\ref{plotqpol} 
at $Q^2 = 10\mbox{GeV}^2$  the PDF when 
$S_q \neq 1$, or when we set abruptly $S_q = 1$, the effect of the
activation function reduces their values at small $x$ because at large
$x$ it becomes close to 1, the most striking effect is observed for the polarized
gluon.

We conclude that the  model of a neural structure for the PDF is perfectly
compatible with unpolarized and polarized experimental data.
\section{Conclusion}
The statistical model provides a better knowledge of the nature of
the parton distribution functions in the sense that their usual properties appears
as a simple consequence of the statistical functions (Fermi or Bose-Eistein)
and the thermodynamical potentials. 
The model
gives a fairly good description of unpolarized and polarized
 experimental data with a reduced number of parameters and also presents
a good laboratory to explore the partons structure.
The sign of the polarized PDF for the quarks is fixed by the potentials and
the dominance of the unpolarized and polarized $u$ over the $d$ appears in
a quite natural way. The calculation of the entropy for the two states $|2u +d>$ 
and $|u+d+s>$ satisfies a maximum entropy principle with the potentials obtained
from the experimental value of the PDF parameters. We have also proven that this
optimum principle is valid for the structure functions $F^2_p$, $g^1_p$ and at the end
to the quarks themselves.

A description of the polarized gluon in term of a neural model gives a more
physical insight on its strucrure and removes the arbitrariness of the original 
formulation. An extension of the neural approach to quarks is derived leading
to a coherent picture of the partons structure which describes both unpolarized
and polarized experimental data. 
From a pure  numerical point of view the polynomial and the statistical approaches
give the same results, however, the last one provides a new explanation of
the parton structure. It is clear that a neuron is not a parton but we have shown
that the mathematical formulation applied to the former can be extended to the later.
This first approach certainly needs further
developments by considering helicity dependent activation functions, and
also an extension to heavy quarks has to be envisaged.\\

\end{document}